\documentclass{PoS}
\usepackage{wrapfig,graphicx}
\usepackage{siunitx} 
\usepackage{subfigure}

\title{Calibrating the photon detection efficiency in IceCube}
\ShortTitle{Calibrating the photon detection efficiency in IceCube} 
\author{\speaker{D.~Tosi} and C.~Wendt for the IceCube Collaboration \\
WIPAC - UW Madison \\
222 West Washington Avenue, suite 500\\
E-mail: \email{delia.tosi@icecube.wisc.edu}, \email{chwendt@icecube.wisc.edu}
}

\abstract{
The IceCube neutrino observatory is composed of more than five thousand light sensors,
Digital Optical Modules (DOMs), installed on the surface and at depths
between 1450 and 2450\,m in clear ice at the South Pole. Each DOM
incorporates a 10-inch diameter photomultiplier tube (PMT) intended to detect
light emitted when high energy neutrinos interact with atoms in the ice. 
Depending on the energy of the neutrino and the distance from secondary
particle tracks, PMTs can be hit by up to several thousand photons within a few hundred nanoseconds. The number of photons per PMT and their time distribution is used to reject
background events and to determine the energy and direction of each
neutrino. 
The detector energy scale was established from previous lab measurements
of DOM optical sensitivity, then refined based on observed light
yield from stopping muons and calibration of ice properties.
A laboratory
setup has now been developed  to more precisely measure the DOM optical
sensitivity as a function of angle and wavelength.  DOMs are calibrated in
water using a broad beam of light whose intensity is measured with a NIST
calibrated photodiode.  This study will refine the current knowledge of the
IceCube response and lay a foundation for future precision upgrades to the
detector.}
\FullConference{Technology and Instrumentation in Particle Physics 2014,\\
		2-6 June, 2014\\
		Amsterdam, the Netherlands}

\begin{document}
\section{Introduction}
\begin{wrapfigure}{r}{0.45\textwidth}
\begin{center}
\vspace{-25pt}
\includegraphics[width=0.45\textwidth]{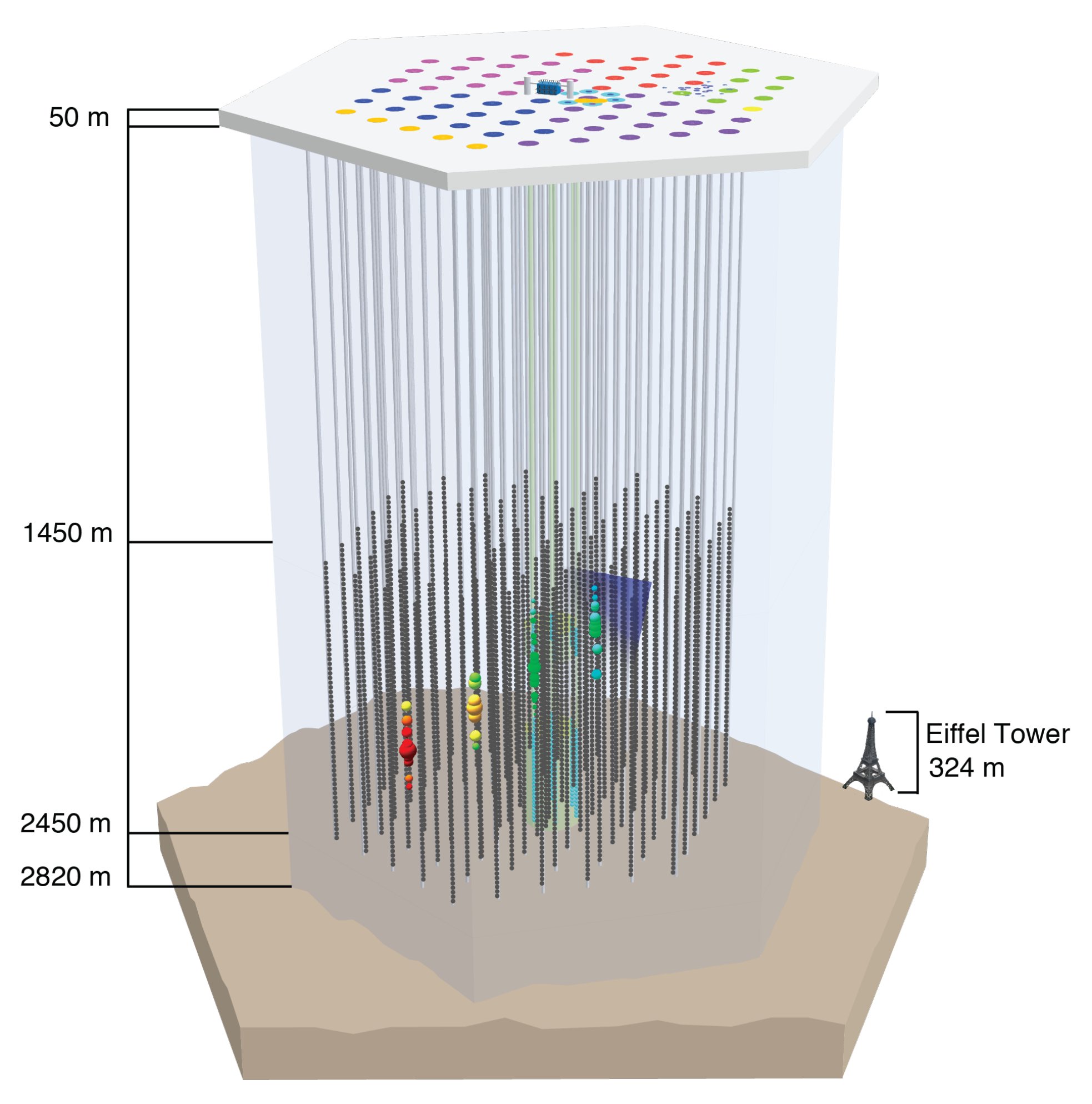} 
\caption{IceCube detector with example event. The colored dots represent DOMs detecting the Cherenkov light from a muon produced by a neutrino interaction.}
\label{fig:IceCube}
\end{center}
\end{wrapfigure}

IceCube is a neutrino telescope installed in the South Pole ice. As pictured in Figure~\ref{fig:IceCube}, the detector is made of 86 strings, mostly spaced 125\,m apart. Each string has 60 Digital Optical Modules (DOMs) deployed at depths between 1450 and 2450\,m, for a total instrumented volume of about 1\,km$^{3}$. A DOM consists of a pressure resistant glass sphere containing a 10-inch diameter R7081-02 photomultiplier tube (PMT) made by Hamamatsu Photonics, and a custom-design single board data acquisition computer. 
Neutrino interactions in ice produce secondary charged particles which are sufficiently energetic to generate Cherenkov light that reaches the DOMs.  Some of the photons striking the PMTs' photocathode sensitive area are detected and amplified, resulting in waveforms that record the amount and timing of arriving light.  The DOMs digitize the PMT waveforms and send them to a counting house at the surface above the center of the array, where waveform information from multiple strings is combined for each interaction into an "event".

Every recorded event is reconstructed to establish the direction of the neutrino and deposited energy. The understanding of the energy scale in IceCube is a very important issue which convolves the 
sensitivity of the PMTs and the properties of the ice, known to be inhomogeneous in the detector volume \cite{IceModel}. Currently the energy scale is based on different sets of measurements performed both in laboratory and in-situ (explained in detail respectively in \cite{PMT} and in \cite{EnergyReco}).

In this paper we describe a new laboratory setup designed and
built at the University of Wisconsin, Madison\footnotemark, in order to perform the absolute
calibration of light sensors. 

\footnotetext{IceCube-HAWC shared laboratory.  We acknowledge notable contributions to the setup by Dan Fiorino, Zig Hampel-Arias and Ian Wisher from the HAWC collaboration.\label{ftn:1}}

\section{Laboratory setup} \label{setup}

The lab setup is designed to measure the single photon detection efficiency of a DOM 
when illuminated in a similar way as from neutrino events in IceCube,
where light typically travels 5-200\,m from its source before possible
detection.
Because the DOM diameter is only 13'', at such distances it can be accurately 
described by its efficiency to detect photons arriving from a particular angle,
averaging over all possible points of arrival at the photocathode.
Such an average is implicit in our measurements which use a uniform light beam from a source 6.4\,m away.  

In IceCube, each DOM has its PMT facing downward, and the sensitivity does not vary significantly with rotation
around this vertical axis.  The sensitivity does depend on the polar angle of illumination relative to the DOM axis,
so the lab setup includes a motorized mounting shaft for inclining the DOM axis relative to a fixed vertical beam.
The DOM is immersed in water in order to closely model reflection and refraction effects occurring at the optical interface where the DOM would be embedded in ice.
Effects of PMT gain, discriminator threshold and electronics calibration
are accounted for by using the same software and procedures as
in standard IceCube operations.

\begin{figure}
\centering
\includegraphics[width=1.0\textwidth]{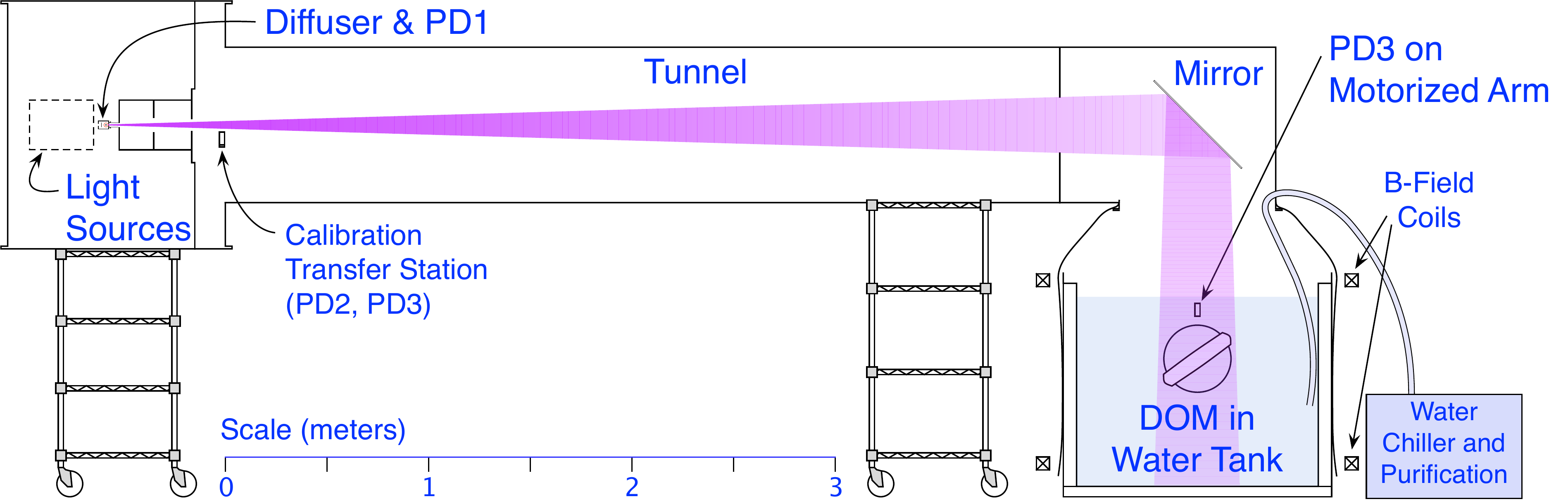}
\caption{Illumination system and water tank.  Photodiodes are placed as shown and used as described in the text:
(PD1) Relative measurement of beam intensity for bright or dim source settings; (PD2) NIST calibrated, used to
calibrate PD3 when temporarily mounted at transfer station; (PD3) Direct measurement of beam flux at DOM for
bright source settings.}
\label{fig:labSetup}
\end{figure}

The illumination system and water tank are sketched in Figure~\ref{fig:labSetup}.
There are several choices of light source, any of which can be directed to
shine into a small diffuser box.  The diffuser box acts as an  integrating sphere, coated inside with white barium sulfate paint and with an optical baffle on the entrance port.
The diffuser's main output port is aimed horizontally towards a mirror 5.3\,m away, which
reflects the beam downward 1.1\,m towards the DOM.  
Photodiode PD1 is mounted on a secondary port of the source diffuser and is 
used to precisely measure changes in beam intensity, including  programmed brightness changes of order $10^5$.
The beam geometry is defined by a series of apertures in the tunnel, creating a uniform circular disc of light with a 16" diameter at the DOM position.
Surfaces in the tunnel and around the DOM are blackened to control stray reflected light.

Currently installed light sources include a pulsed diode laser (405\,nm), pulsed LEDs (400\,nm),
continuous LEDs (370\,nm, 400\,nm, 450\,nm), and lamp with monochromator (320-700\,nm).
Beam intensity entering the diffuser box can be varied over a wide range by means of 
LED current, repetition rate of pulsed sources, neutral density optical filters,
and a simple shutter.

Photodiode PD3 is in a waterproof housing and mounted on a robotic arm, so it can be moved 
into the water above the DOM.  When sources are operated in a bright mode, PD3 is
used to precisely measure the light reaching the DOM, as well as to test uniformity
and boundaries of the beam cross section.  When sources are switched to their dim mode,
PD3 is moved out of the way and PD1 measures the intensity ratio relative to the bright mode
(section \ref{method}).

Closer to the diffuser output, a calibration transfer station allows temporary mounting
of PD3 side by side with a NIST calibrated photodiode, which has a response
known to better than 0.5\% when
illuminated near its center at wavelengths 300\,--\,\SI{700}{\nano\meter}.
For this cross calibration of PD3, an aperture is inserted in front of the transfer station,
reducing the beam cross section to a small spot that can be positioned on center or scanned across the two photodiodes. 
PD1 is used to correct for small variations of source intensity
during the scan.  The calibrated response of PD3 averaged
over its active area is then converted to a value appropriate to a uniform
beam.

An optical fiber pickup can also be moved into the beam at the calibration transfer station,
with the fiber routed to a separate monochromator and PMT.  This allows measurement of the
wavelength spectrum of the beam for any source configuration.

\begin{figure}
\centering
\includegraphics[width=0.8\textwidth]{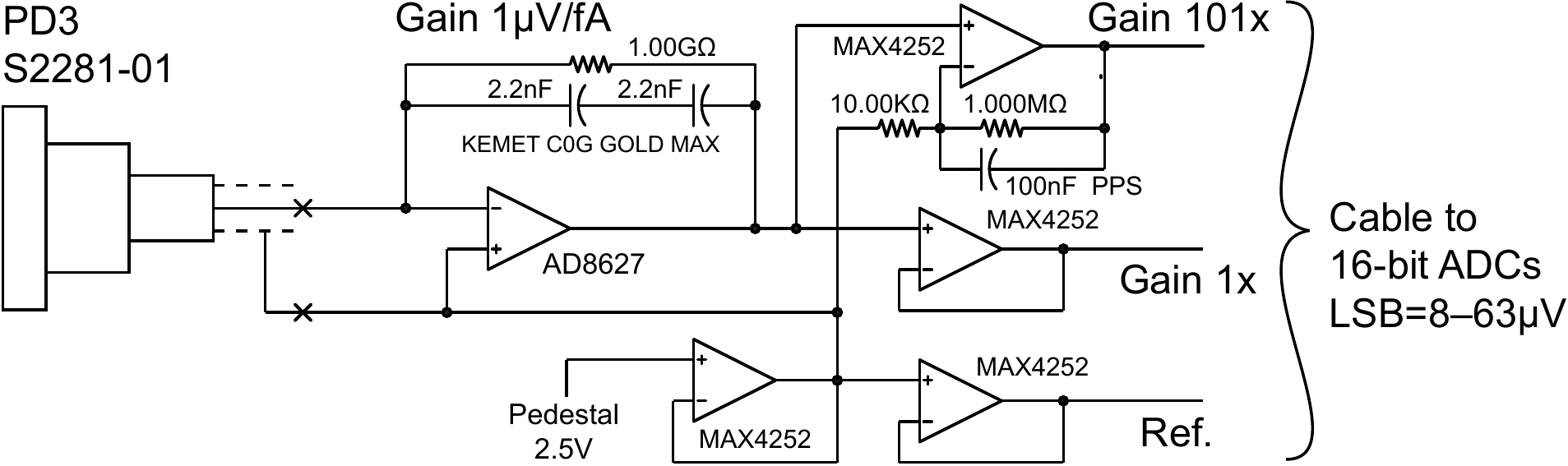}
\caption{Detail of the preamplifier circuit used for photodiode PD3.}
\label{fig:schematic}
\end{figure}

Photodiode currents are measured with custom preamplifier and ADC boards.
Figure~\ref{fig:schematic} shows the PD3 preamplifier, which is built on a 1''-diameter
circuit board and mounted directly on the PD3 connector before encapsulation in a
waterproof housing.  The photodiode current is amplified with gain 
\SI[per-mode=symbol]{1}{\micro\volt\per\femto\ampere}
and sent to an ADC board in the light sources box via a shielded twisted pair cable, 
along with another copy of the signal amplified $\times101$.
Low noise and offset of the AD8627 operational amplifier, 
together with the high shunt impedance
of the photodiode, allows measurement of currents
around \SI{100}{\femto\ampere} 
with good precision when an on-off technique is employed for
subtracting baseline offsets.
The 16-bit ADCs (ADS1110) are configured with
least significant bit 8\,--\,\SI{63}{\micro\volt}.
A similar circuit is used for PD1 and the NIST photodiode 
(Hamamatsu S2281), 
except that relays are provided to switch the first stage between
gains in the range 
\SI[per-mode=symbol]{10}{\micro\volt\per\nano\ampere}
to \SI[per-mode=symbol]{1}{\micro\volt\per\femto\ampere}.

The tank water is continuously pumped through an external chiller and water purification
system.  Maintaining the temperature around $5^\circ$C
avoids introducing high humidity to the tunnel
and source region, where it might affect electronic or optical elements.  The low temperature
also reduces the DOM dark noise rate and discourages bacterial growth in the water.
Wetted mechanical elements are made of stainless steel, glass, anodized aluminum, or
plastics compatible with high purity deionized water.

Surrounding the water tank, Helmholtz coils are arranged for separate control of the
magnetic field in each direction.  These coils cancel the ambient field in the lab and
create a field relative to the DOM axis (in its rotated position) which matches that
experienced at the South Pole.

A blackout curtain hangs from the open end of the tunnel and is secured
around the water tank by a belt.  For replacing the DOM, the curtain is raised and
the tunnel assembly is wheeled aside.  

All system elements are controlled and monitored by Python scripts running on a standard
IceCube data acquisition computer (DOMHub), which also embodies the interface for 
control and readout of the DOM\cite{DOMDAQ}.  
A separate DOM-style data acquisition board is connected to the DOMHub for
recording electronic synchronization pulses when using the laser or pulsed LEDs, 
allowing us to reject out-of-time DOM hits in data analysis.

 \section{Measurements}\label{method}

\begin{table}
\centering
\setlength\extrarowheight{2pt}
\begin{tabular}[c]{|l|c|c|c|c|}
\hline  
Source Setting & Photon flux & PD1 & PD3 & DOM \\
\hline  
Bright & $\rm{10^6/cm^2/sec}$ & 50\,nA & 100\,fA & saturation\\
Dim & $\rm{10/cm^2/sec}$ & 500\,fA & too low & 100-1000\,Hz\\		
\hline  
\end{tabular}
\caption{Approximate response of PD1, PD3 and DOM for bright and dim beams.}\label{tab1}
\end{table} 

Our fundamental reference for measuring light levels is a NIST calibrated photodiode \cite{NIST}.
As explained in section~\ref{setup}, we use this precision photodiode to calibrate PD3 
which has been encapsulated for underwater use.  
We then install PD3 in the water tank and turn on one of the light sources,
resulting in a PD3 current around
\SI{100}{\femto\ampere}, measured to within 0.5\%.
This gives a direct measurement of beam flux in the tank, typically 
$10^6$\,photons/cm$^2$/sec at high brightness settings.

However, the PD3 current signal cannot be precisely measured at the much lower brightness settings needed for DOM sensitivity calibration.
For beam fluxes below
100\,photons/cm$^2$/sec,
we instead rely on PD1 which is much closer to the source.  
The larger PD1 signal is also proportional to the beam flux, but with an unknown scale factor that must be calibrated against the direct PD3 measurement.  For this purpose we perform a bright PD3 measurement simultaneously with a PD1 measurement, yielding the beam flux scale factor:
$$\rm{Flux}=0.0226\rm{\,\,photons/cm^2/sec}\cdot(\rm{PD1\,\,current\,/\,fA})$$
Since PD1 and PD3 are both observing outputs of the source diffuser box, this scale factor applies equally when the source is operated in dim mode.  
The switchable gain preamp used with PD1 facilitates measurement of both bright and dim beams with good precision (Table 1).  For example, a dim beam with 10~photons/cm$^2$/sec at 400\,nm gives
DOM count rates up to 1000\,Hz (depending on orientation), with PD1 currents around 500\,fA.
Such dim measurements are then made for each polar angle and wavelength and
used to calculate the final DOM photon sensitivity.

\begin{figure}[ht]
\begin{minipage}[b]{0.47\linewidth}
\centering
\includegraphics[width=\textwidth,height=1\textwidth]{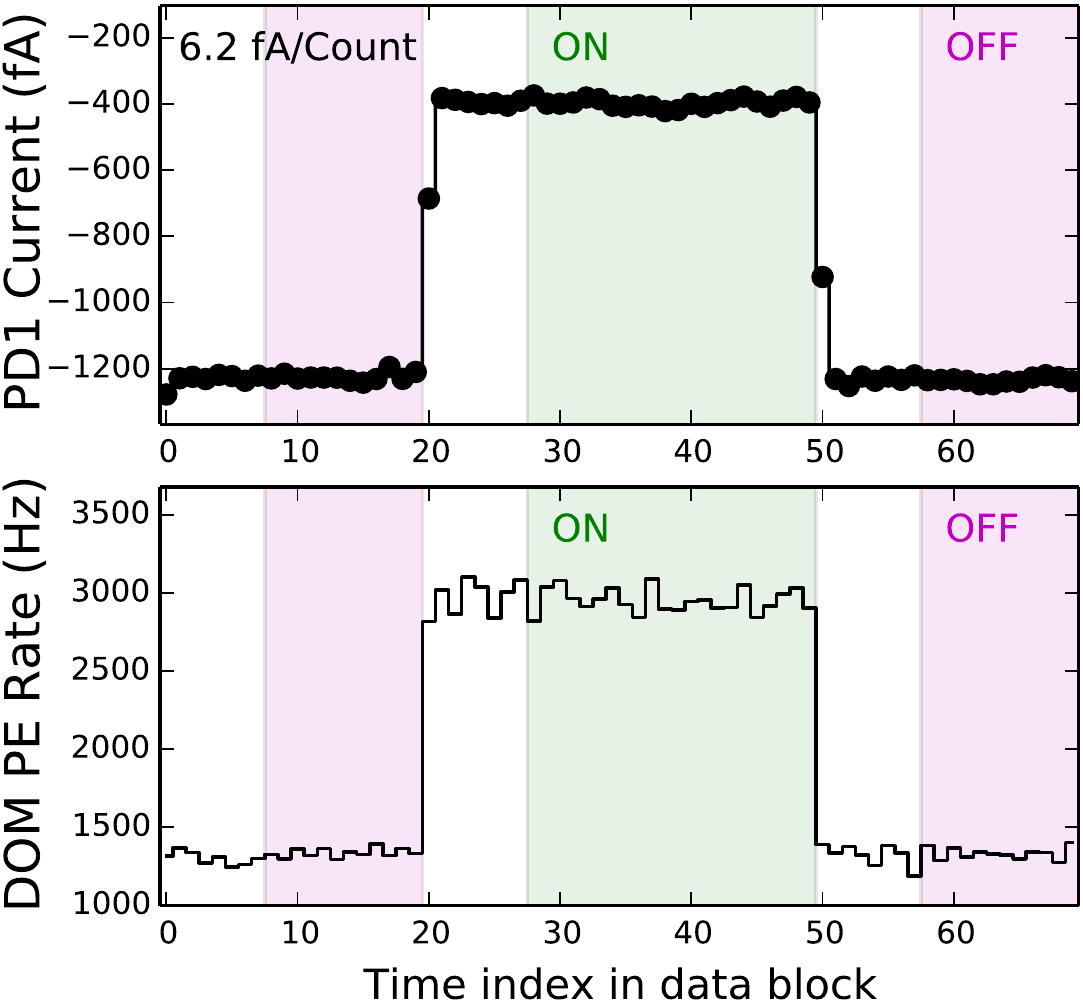}
\caption{\label{fig:signals}
Example signals from PD1 and DOM using steady LED light at 400nm.  ''OFF'' intervals reveal
offsets to be subtracted in ''ON'' intervals.
}
\end{minipage}
\hspace{0.02\linewidth}
\begin{minipage}[b]{0.47\linewidth}
\centering
\includegraphics[width=\textwidth,height=1\textwidth]{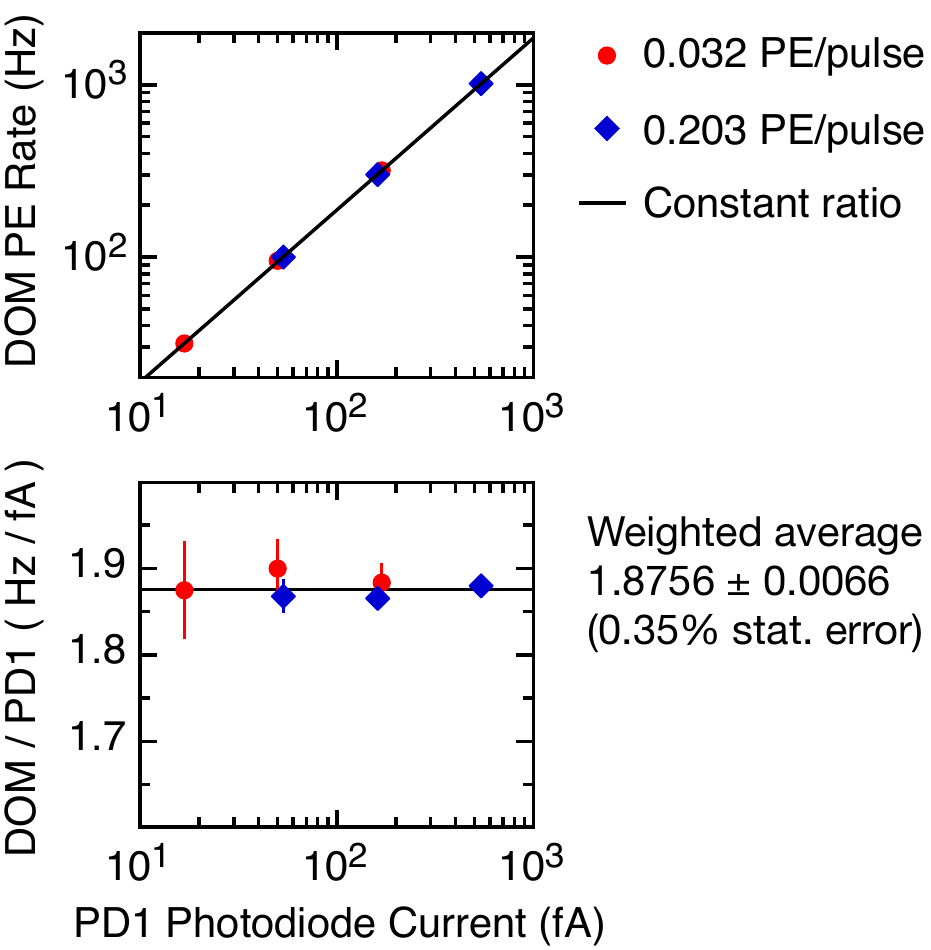}
\caption{\label{fig:scaling}Comparison of DOM photon detection rate with PD1 current, using
pulsed laser light at 405\,nm.}
\end{minipage}
\end{figure}

In more detail, Figure~\ref{fig:signals} shows typical signals observed in PD1 and the DOM
when dimly illuminated with steady LED light at 400\,nm.  Using the relation between PD1 current
and photon flux in the tank, the ratio of these signals leads directly to the DOM's photon
detection efficiency.
Each measurement block provides an average over 20 seconds with source on, 
and generally requires 
subtraction of a baseline value determined during source-off periods before and after.
With source off, the PD1 current readout is nonzero due to slowly drifting electronic
offsets  (possibly negative as in Figure~\ref{fig:signals}), 
whereas the DOM shows its PMT dark noise rate.  
Statistical fluctuations are observed to
be quite small, and can be reduced further by averaging results from multiple such data
blocks.
The single photoelectron (SPE) count rate in the DOM is defined by discriminator triggers 
above a threshold of 0.25 times the peak charge observed in the SPE charge spectrum,
which is the same condition used to count photons in the IceCube detector.  In order
to avoid counting afterpulses as part of the detection efficiency, each discriminator trigger starts
a programmed dead time of \SI{12.8}{\micro\second}, and the remaining live time is
used to calculate the corrected SPE rate.

Figure~\ref{fig:scaling} shows that similar results can be obtained from a pulsed light
source.  In this case the DOM was programmed to report individual PMT hit times and only hits
coincident with light pulses were counted in the analysis.  
While the time coincidence mostly avoids the effects of PMT dark noise
and afterpulses, a statistical correction factor has to be applied when individual pulses are 
bright enough that often two photons are detected simultaneously without being resolved (blue diamond data points).
The figure shows that proper scaling between DOM counts and PD1 current is maintained 
while changing the per-pulse brightness or the overall rate.
With pulse rate 5\,kHz and DOM occupancy 0.203 photoelectrons per pulse, 
the ratio is measured with statistical precision better
than 0.5\% by averaging 3 time blocks similar to Figure~\ref{fig:signals}.

\section{Photodiode linearity} \label{nonlinearity}
As explained in the previous section, the DOM measurement is done at much lower light level than  can be directly measured with PD3 mounted near the DOM in the water tank.
The calibration and subsequent use of PD1 for measuring photon fluxes involves a range of photodiode current spanning over five orders of magnitude, as indicated by the ratio between the photodiode currents shown in Table \ref{tab1}. 
Therefore, we need to check the linearity of the photodiode and its readout over this range.
The Hamamatsu S2281 is known for its linearity \cite{NIST,Lei1993}, but we still need to test for non-ideal behavior in our specific application, which includes the amplifier and readout chain. 

In order to detect small nonlinear effects, we study the measured response to two steady LEDs (at 400\,nm) with
similar brightness.
We calculate the ratio of current measured
with both LEDs turned on to the sum of currents measured when each LED is turned on by itself.  
Any departure of this ratio from unity indicates accumulation of error from nonlinearity over a factor of two in 
brightness delivered to the photodiode.  The accumulated nonlinearity between two very different brightnesses is then
approximately the sum of contributions from each factor two between those brightnesses (assuming only small
deviations), so we need to measure each contribution very precisely.  This method does not require
any particular linearity in the source brightness control, only that sources have stable and reproducible output when
enabled. 

The light from both LEDs is arranged to shine into the source diffuser entrance port, so a given measurement
cycle requires only changes to the LED drive currents and no changes in optical paths.
Each LED is driven independently by a current source which can be
programmed by a 16-bit DAC and a bright/dim range setting, yielding very stable light outputs 
spanning more than four orders of magnitude.
In order to avoid any effect on one LED's brightness when the other LED is switched on, 
each LED is mounted with its DAC control on a separate circuit board with
individual voltage regulation and digital communications to a host microcontroller.
Generally we use the same brightness DAC settings for both LEDs in a given measurement. 
To reach the lowest light levels of interest, we insert an optical attenuator rather than reducing the LED drive current
so much that noise becomes an issue.

\begin{wrapfigure}{ht}{0.55\textwidth}
\vspace{-10pt}
\centering
\includegraphics[width=0.5\textwidth]{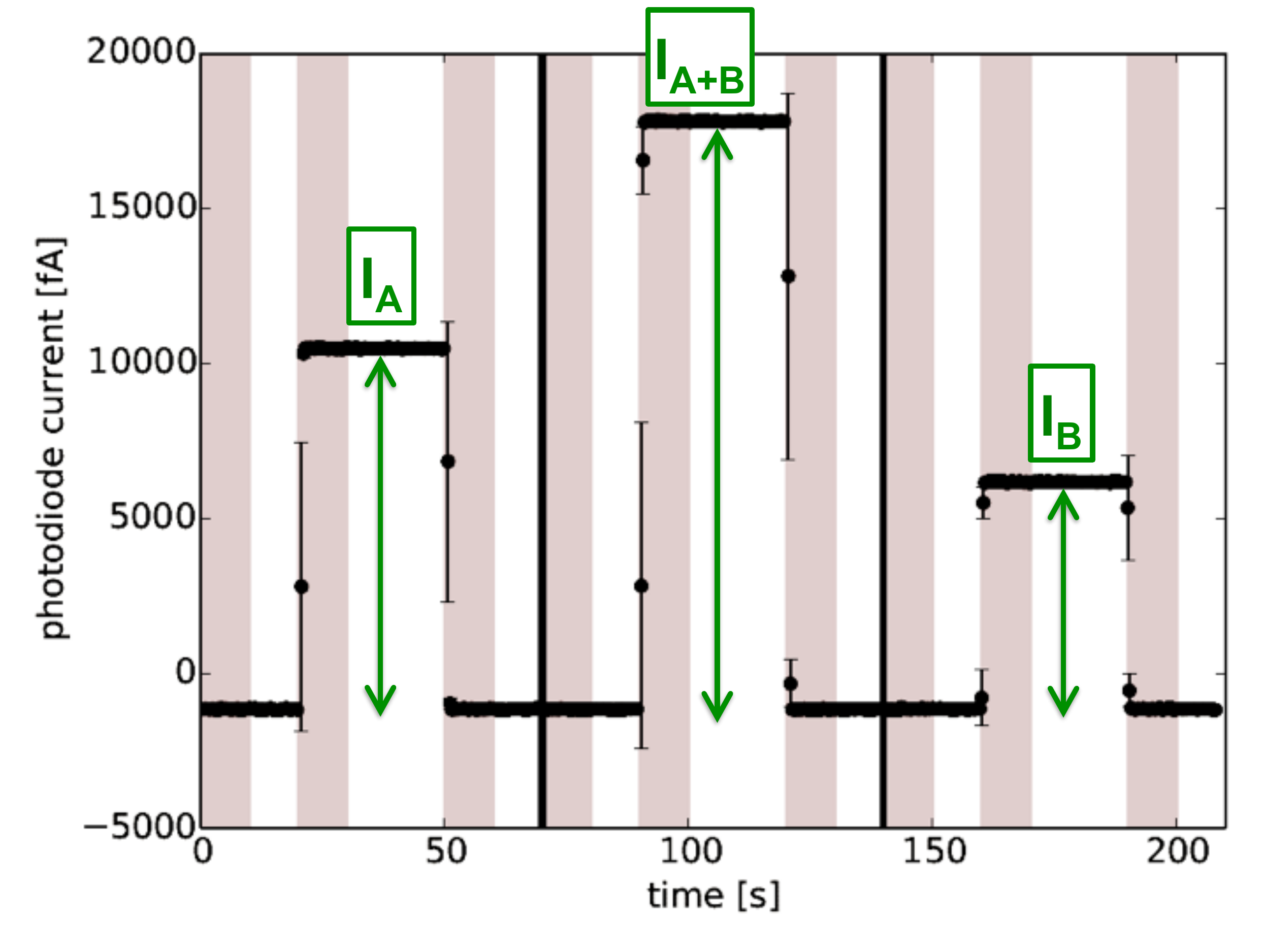}
\vspace{-10pt}
\caption{\label{fig:linearityMethod}Illustration of the method used to study the photodiode linearity. The black vertical lines indicate the three segments (see text); the red bands show the transient time which is excluded from the analysis.}
\end{wrapfigure}

Figure \ref{fig:linearityMethod} illustrates the method in detail.  With both LEDs off, there is an offset that must
be measured and subtracted.
We first measure the increment in current when LED-A only is turned on (segment 1); then the same with both LED-A and LED-B on (segment 2), and finally with only LED-B on (segment 3). 
For every segment we have an off-time of 10\,s, an on-time of 20\,s, and every time the source switches on or off, we record but do not use data for a transient time of 10\,s. One full segment therefore lasts 70\,s, one cycle (three segments) lasts 210\,s; we record at least 30 cycles for every light setting and compute average ratios as described earlier.

\begin{figure}[t]
\centering
\subfigure[]{
\includegraphics[trim=20 20 0 0,clip,width=0.475\linewidth,height = 0.41\linewidth]{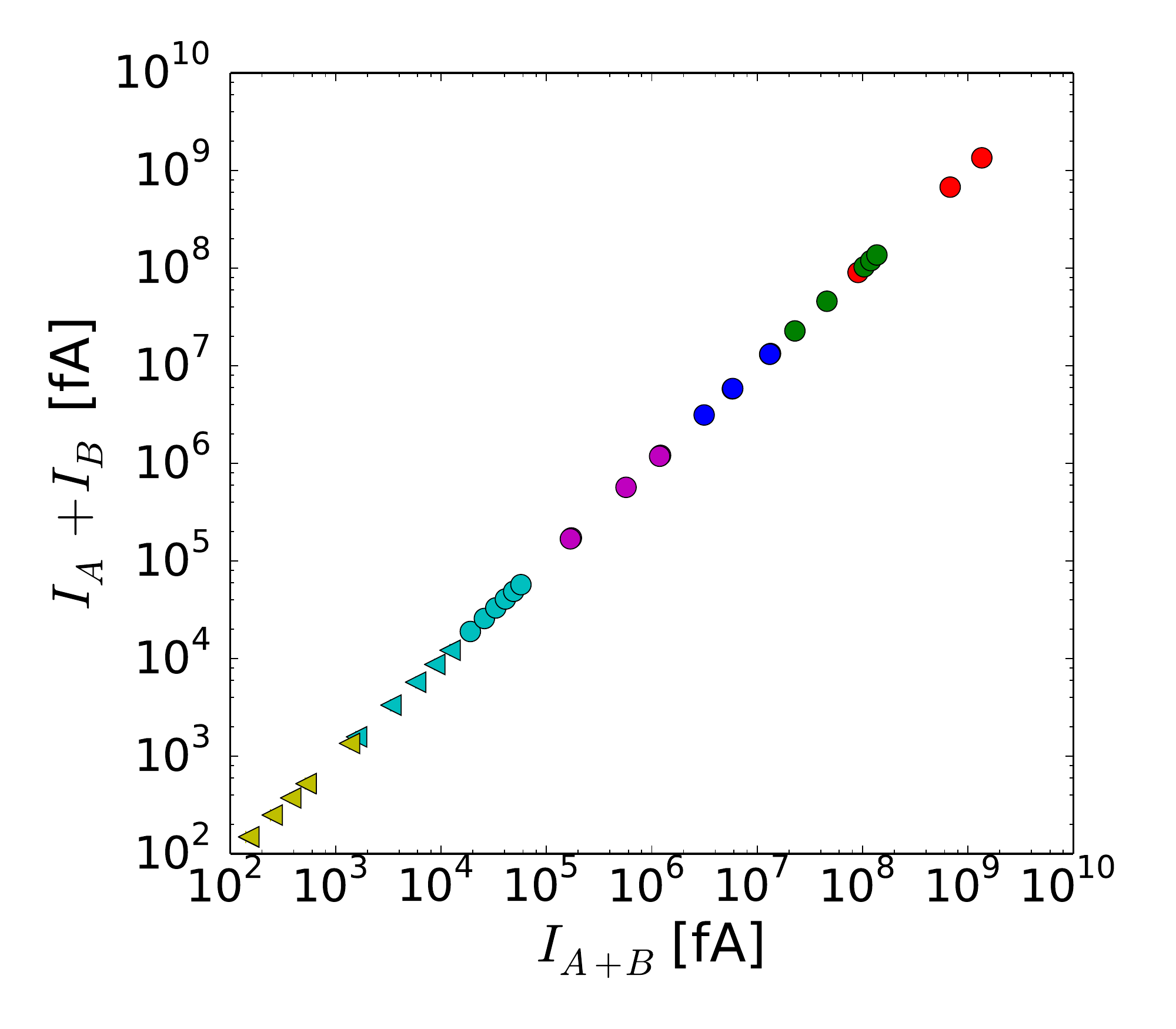}
\vspace{-50pt}
\label{fig:linearityCurrents}
}
\hspace{-20pt}
\subfigure[] {
\includegraphics[trim=20 20 0 0,clip,width=0.505\linewidth,height = 0.41\linewidth]{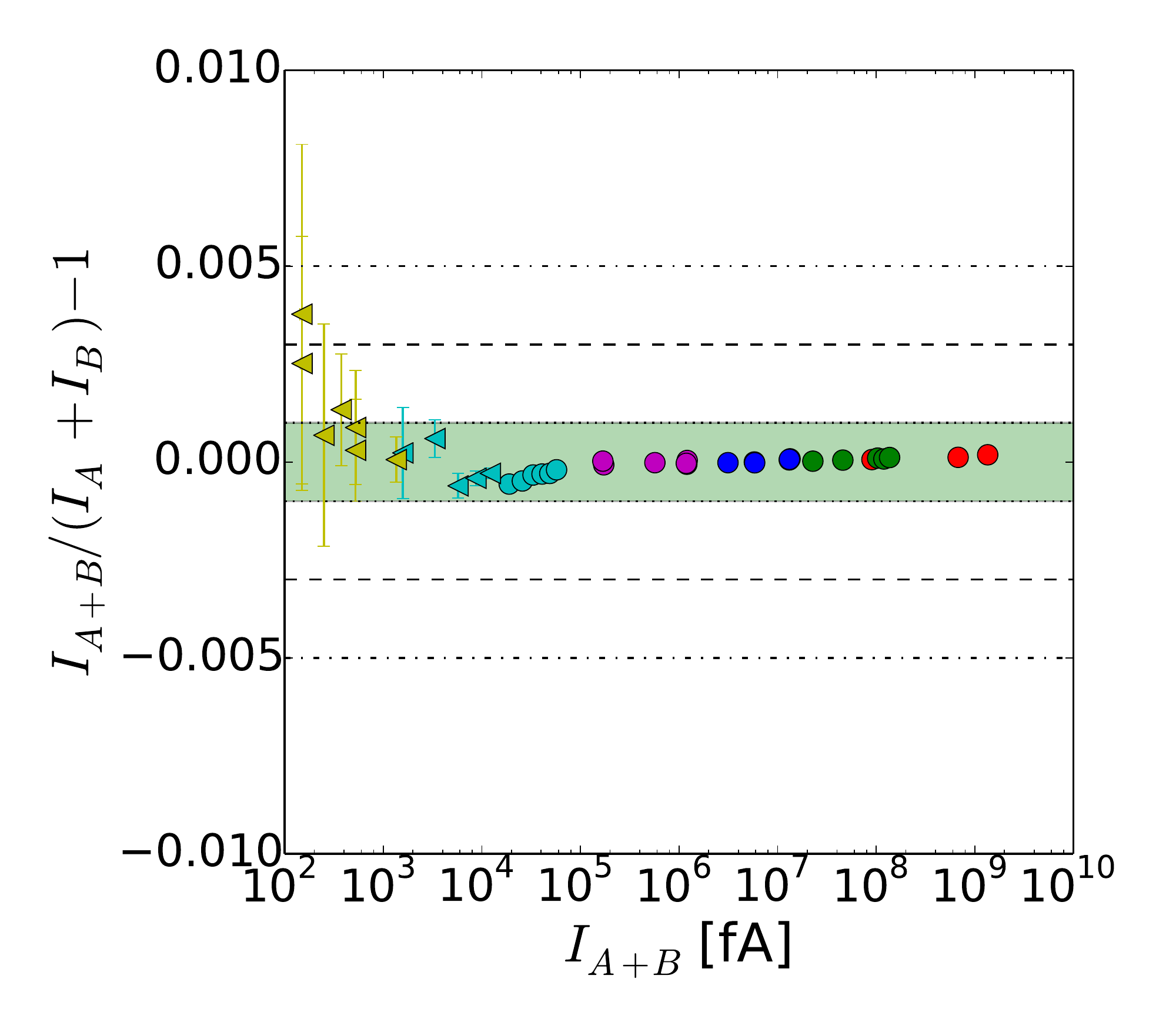}
\vspace{-50pt}
\label{fig:linearityRatio}
}
\vspace{-10pt}
\caption{
\protect\subref{fig:linearityCurrents} Linearity in PD1 response as measured over light levels spanning nine orders of magnitude in current.
\protect\subref{fig:linearityRatio} Nonlinearity at different light levels for PD1. Each point describes the nonlinearity for one doubling of intensity at a specific value of the photon flux. The green band shows the $\pm$0.1\% level.}
\label{fig:lin}
\end{figure}

The result of the linearity measurements as a function of the photodiode current is shown in Figure~\ref{fig:linearityCurrents}. The deviation of the ratio from unity is plotted in Figure~\ref{fig:linearityRatio}. 
Generally the measurements obey linearity within uncertainties well below 0.1\%; statistical uncertainties and subtle
effects at very dim light levels are currently the limiting factor, but still contributing only 0.1-0.2\%.
Considering the sum of 
nonlinearities from each doubling of intensity from \SI{500}{\femto\ampere} to \SI{50}{\nano\ampere}, we
expect the total nonlinearity error to be $\leq$0.2\% when comparing dim and bright light levels with PD1.
There is also some uncertainty in the relative gains of the PD1 preamplifier which need to be chosen for the different
light levels; theoretically those are set by high precision resistors but their ratios can also be measured
with precision comparable to the nonlinearity results using the same LED sources.

\section{Summary}
A lab setup has been developed to measure the DOM absolute optical sensitivity 
as a function of incident polar angle and wavelength. An optical system illuminates the DOM uniformly
from a certain direction, with several pulsed and steady light sources available for wavelength selection and systematic studies. The calibration is transferred from a NIST calibrated photodiode to the DOM by using additional photodiodes which monitor the beam, as its intensity is varied over five orders of magnitude. We verified that the error introduced by nonlinearities of the system, operated at the brightest and dimmest limit of the dynamic range, is less than 0.2\%.
In the near future, we plan to calibrate 10 to 20 of about 100 leftover DOMs, and we foresee each measurement to take about one week. Since relative sensitivities of all IceCube DOMs were measured in pre-deployment testing, a DOM dependent correction factor can be used to improve the understanding of the IceCube detector and the energy scale.
The calibration system will be also available for calibration of optical sensors to be used in HAWC\footnotemark[\value{footnote}] and future in-ice detectors such as PINGU \cite{PINGU}.


\end{document}